\documentclass[a4paper]{jpconf}
\usepackage{graphicx,hyperref}

\begin{document}

\title{HEPData: a repository for high energy physics data}

\author{Eamonn Maguire$^1$, Lukas Heinrich$^2$ and Graeme Watt$^3$}
\address{$^1$ CERN, Geneva, Switzerland}
\address{$^2$ Department of Physics, New York University, New York, USA}
\address{$^3$ IPPP, Department of Physics, Durham University, Durham, UK}

\ead{info@hepdata.net}

\begin{abstract}
  The Durham High Energy Physics Database (HEPData) has been built up over the
past four decades as a unique open-access repository for scattering data from
experimental particle physics papers.  It comprises data points underlying
several thousand publications.  Over the last two years, the HEPData software
has been completely rewritten using modern computing technologies as an overlay
on the Invenio~v3 digital library framework.  The software is open source with
the new site available at \url{https://hepdata.net} now replacing the previous
site at \url{http://hepdata.cedar.ac.uk}.  In this write-up, we describe the
development of the new site and explain some of the advantages it offers over
the previous platform.
\end{abstract}

\section{Introduction}

The Durham High Energy Physics Database (HEPData), a unique open-access
repository for scattering data from experimental particle physics papers, has a
long history dating back to the 1970s.  It currently comprises data related to
several thousand publications including those from the Large Hadron Collider
(LHC).  These are generally the numbers corresponding to the data points
either plotted or tabulated in the publications, ``Level 1'' according to the
DPHEP~\cite{Mount:2009aa} classification, and HEPData is therefore
complementary to the recent CERN Open Data Portal
(\url{http://opendata.cern.ch}) which focuses on the release of data from
Levels 2 and 3.  The traditional focus of HEPData has been on measurements such
as production cross sections and so the domain differs from the compilation of
particle properties provided by the Particle Data Group
(\url{http://www-pdg.lbl.gov}).  In recent years HEPData has expanded beyond
the traditional (unfolded and background-subtracted) measurements to also
include data relevant for ``recasting'' LHC searches for physics beyond the
Standard Model.  The scope of HEPData is also being broadened to include
data from particle decays and neutrino experiments, and potentially low-energy
data relevant for tuning of the Geant4 detector simulation toolkit.

The HEPData project last underwent a major redevelopment around a decade
ago~\cite{Buckley:2010jn}, as part of the work of the CEDAR
collaboration~\cite{Buckley:2007hi}, where data was migrated from a legacy
hierarchical database to a modern relational database (MySQL) and a web
interface built on CGI scripts was replaced by a Java-based web interface.  The
old HepData site (\url{http://hepdata.cedar.ac.uk}) ran on a single machine
hosted at the Institute for Particle Physics Phenomenology (IPPP) at Durham
University.  Over the last two years, a complete rewrite has once again been
undertaken to use more modern computing technologies.  The new site
(\url{https://hepdata.net}) is hosted on a number of machines provided by CERN
OpenStack and offers several advantages and new features compared to the old
site.  In this write-up, we describe the development of the new HEPData site
(note the slight rebranding denoted by the change in capitalisation of
``HEPData'' compared to the old ``HepData'').  The code is open source and
available from a dedicated GitHub organisation
(\url{https://github.com/HEPData}).

\section{Migration}

The old HepData site stored all information in a (MySQL) database.  To add a
new record, all data needed to be first manually transformed into a standard
``input'' text format, consisting of metadata for each table followed by data
points in a structured format.  The uploaded input file was then parsed by a
Perl script to insert the information into the database.  For the new HEPData
site, we decided to store records as text files rather than in a database,
since only the metadata (and not the actual data points) needs to be made
searchable.  Rather than retaining the old \emph{ad hoc} ``input'' text format,
we defined a new text format
(\url{http://github.com/HEPData/hepdata-submission}) using YAML
(\url{http://yaml.org}), which is a superset of JSON that is more
human-readable.  We
investigated the possibility of a new universal input format using ROOT
(\url{https://root.cern.ch}), but it was not suited to representing all of the
diverse data types already present in HepData, particularly the metadata that
describe each of the tables.  All data from the existing HepData database was
then exported to the new YAML format for migration to the new system.  A
validator (\url{http://github.com/HEPData/hepdata-validator}) was also written
to ensure that the YAML/JSON files conform to the defined
\href{https://github.com/HEPData/hepdata-validator/tree/master/hepdata_validator/schemas}{JSON schema}.
 Migrated YAML
files (and future submitted files) are stored on the CERN EOS file system.

\section{Software}

\begin{figure}
  \begin{center}
    \frame{\includegraphics[width=\textwidth]{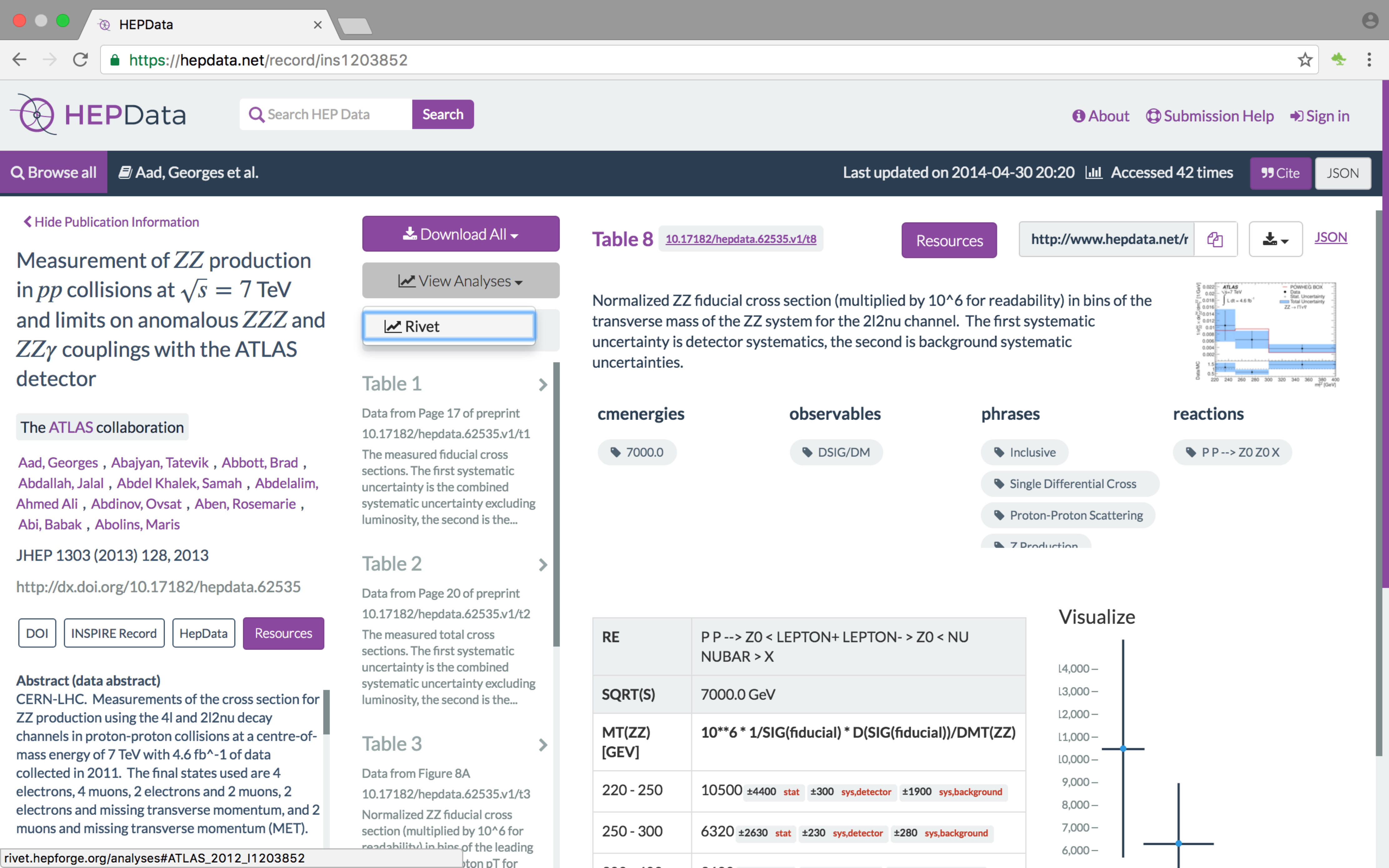}}
  \end{center}
  \caption{\label{fig:record}Screenshot of a typical record on the new HEPData site.}
\end{figure}

The HEPData software was rewritten from the ground up, predominantly in the
Python and JavaScript programming languages, as an overlay on the Invenio~v3
(\url{http://inveniosoftware.org}) digital library framework, but with a very
large degree of customisation.  A screenshot of a typical data record is shown
in figure~\ref{fig:record}.
As with the previous HepData site, the main added value of storing the data in
a standard format is that the data points can be automatically converted to
various formats (see section~\ref{sec:conversion}) and visualised.  Tables and
scatter plots, or heatmap plots if there is more than one independent variable,
are rendered with custom JavaScript code making use of the D3.js library
(\url{https://d3js.org}), with options to switch on and off various elements.
 Auxiliary files, such as the original ATLAS plot shown in the top-right of
figure~\ref{fig:record}, can be attached or linked to either individual tables
or the whole record.  A semi-automated way was developed to make links to
analysis code within frameworks such as
\href{https://rivet.hepforge.org}{Rivet}~\cite{Buckley:2010ar} (see middle
panel of figure~\ref{fig:record}), where the framework authors provide a JSON
file listing available analyses.

\section{Discoverability}

\begin{figure}
  \begin{center}
    \frame{\includegraphics[width=\textwidth]{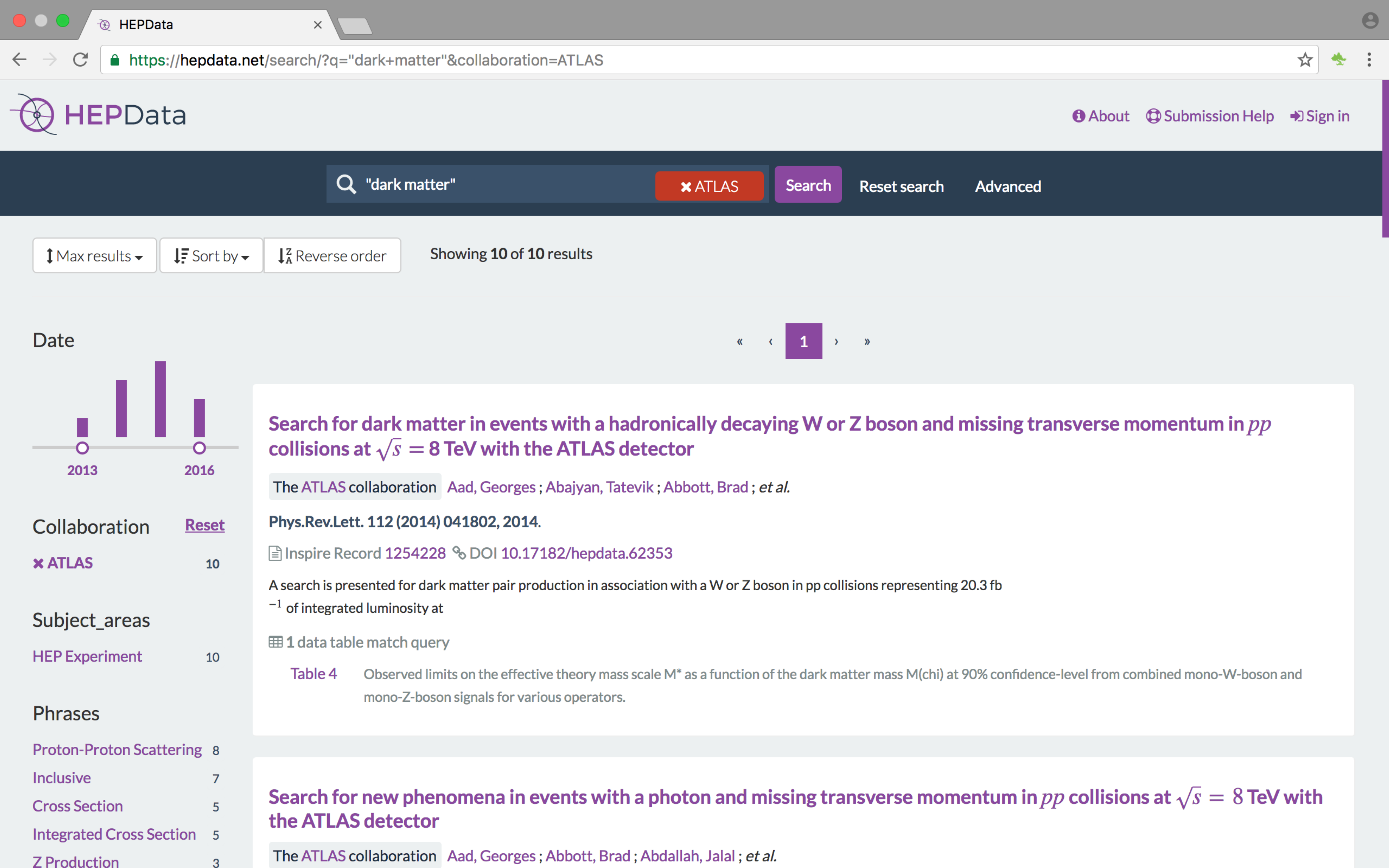}}
  \end{center}
  \caption{\label{fig:search}Screenshot of a typical search on the new HEPData
site.}
\end{figure}

\begin{figure}
  \begin{center}
    \frame{\includegraphics[width=\textwidth]{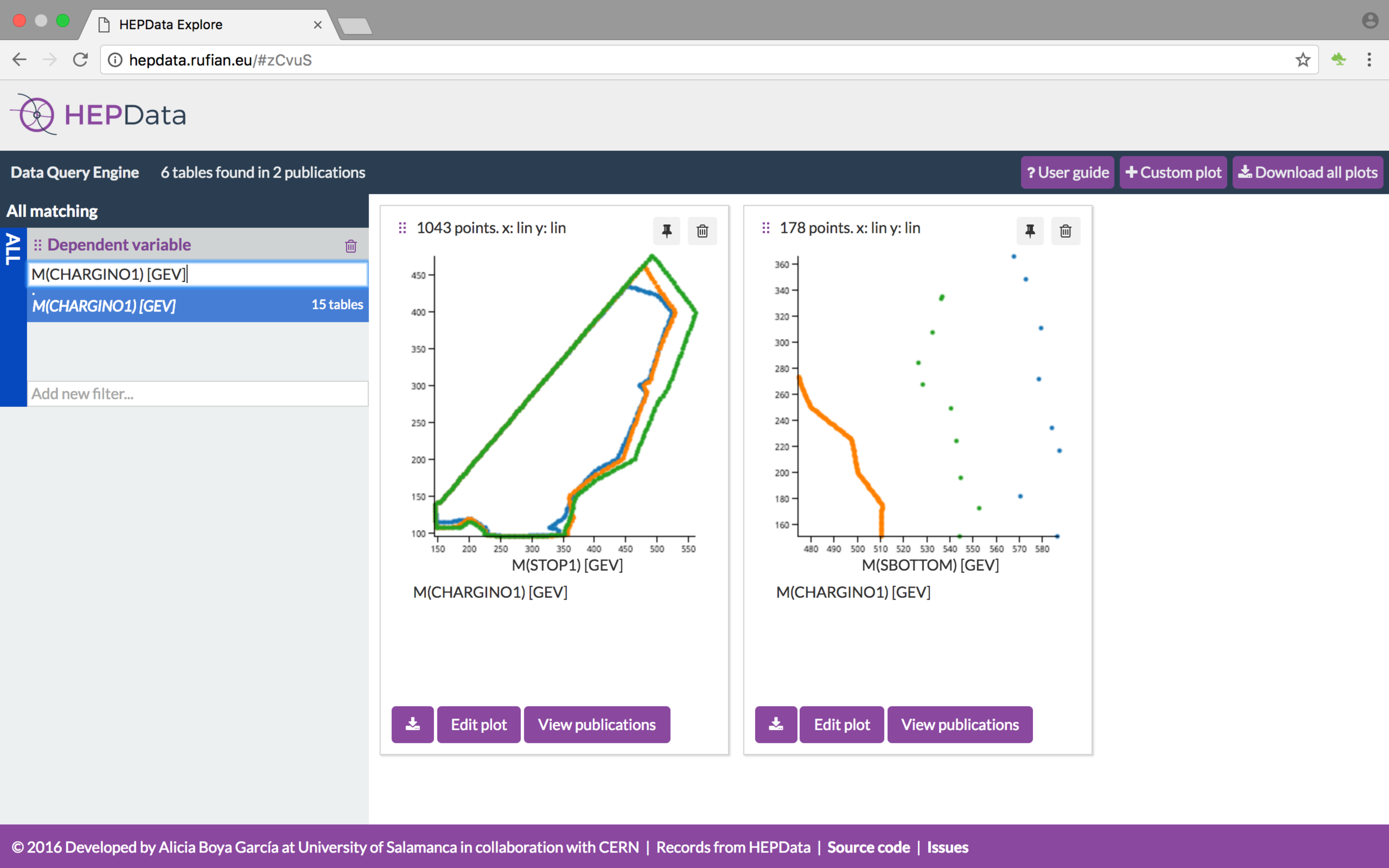}}
  \end{center}
  \caption{\label{fig:explore}Screenshot of an alternative data-driven query
module for HEPData.}
\end{figure}

The new HEPData software uses a PostgreSQL database (with a new data model
compared to the previous MySQL database), indexed with
\href{https://www.elastic.co}{Elasticsearch} to provide fast and powerful
searching across all metadata fields.  A screenshot of a typical search is
shown in figure~\ref{fig:search}.
More specific searches on keywords such as \verb+cmenergies+,
\verb+observables+, \verb+phrases+ and \verb+reactions+ are also possible.
 Faceted search is implemented for certain fields; see the left-hand panel in
figure~\ref{fig:search}.  All content in HEPData is semantically enriched using
\href{https://schema.org/}{\emph{Schema.org}} vocabulary.  This means that
Google and other search engines know more about the content, and that the
content can be automatically retrieved and interpreted by developers.  An
alternative data-driven search module for HEPData has been developed by Alicia
Boya Garc\'{i}a for a Master's project at the University of Salamanca and a
prototype is available at \url{http://hepdata.rufian.eu}; see
figure~\ref{fig:explore}.
This
\href{https://github.com/HEPData/hepdata-explore}{\texttt{hepdata-explore}}
package offers the possibility for a user to filter data from multiple
publications, producing composite plots, with the ability to download numerical
values.

In 2012, a first attempt was made to integrate the old HepData site with the
Inspire HEP literature database (\url{http://inspirehep.net}) by harvesting the
HepData tables and creating new Inspire records for each table.  This first
attempt was an important step forward, but the integration was never fully
completed.  The Inspire service is also being rewritten using Invenio~v3 and so
integration with the new HEPData site should be straightforward.  The new
HEPData records are now versioned in a way similar to
\href{https://arxiv.org/}{arXiv} papers, meaning that mistakes in an original
submission can be corrected with a ``version 2'' and the original ``version 1''
is retained and not overwritten.  \href{https://www.doi.org}{Digital object
identifiers (DOIs)} are minted separately for each version via
\href{https://www.datacite.org}{DataCite} for both whole records and
individual tables.  Eventually, the DOIs will allow citation of HEPData
records to be tracked by Inspire in a similar way to publications.

\section{Conversion} \label{sec:conversion}

\begin{figure}
  \begin{center}
    \includegraphics[width=0.5\textwidth]{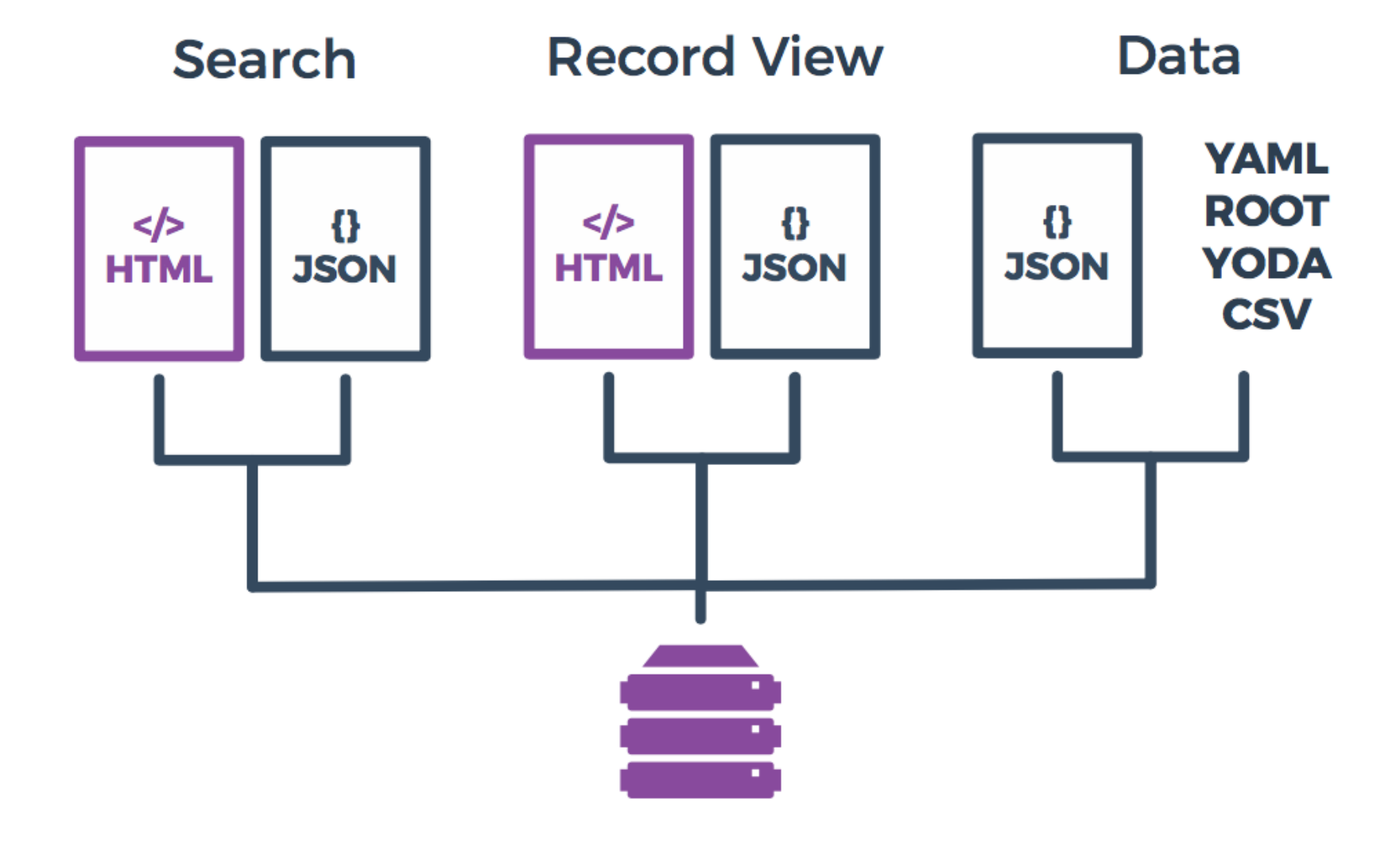}
    \hspace{2pc}%
    \begin{minipage}[b]{14pc}
      \caption{\label{fig:programmatic}Diagram illustrating programmatic access
and export formats.  Search results, record display, and data tables can all be
obtained in a JSON format.  Data tables can be exported in either the native
YAML format, or converted to alternative CSV, ROOT and YODA formats.}
    \end{minipage}
  \end{center}
\end{figure}

The HTML web pages for search results and record display on the new HEPData
site all have a JSON equivalent; see figure~\ref{fig:programmatic}.
The JSON format allows programmatic access, from scripts or applications such
as Mathematica, simply by adding the option \texttt{format=json} to the URL.
 Conversion to various export formats is provided by a separate extensible
package (\url{http://github.com/HEPData/hepdata-converter}), originally
developed by a CERN summer student, Micha\l{} Szostak~\cite{Szostak:2055193}.
Current output formats are listed below:
\begin{description}
\item [\href{http://yaml.org}{YAML}] The native HEPData format of the data
tables (see \href{http://github.com/HEPData/hepdata-submission}{\texttt{hepdata-submission}}).
\item [\href{https://en.wikipedia.org/wiki/Comma-separated_values}{CSV}] A
simple text format consisting of comma-separated values, which can be imported
into many widely-used applications such as an Excel spreadsheet.
\item [\href{https://root.cern.ch}{ROOT}] A binary \texttt{.root} file rather
than a \href{https://root.cern.ch/cint}{CINT} script, with each table in a
separate directory.  For numeric data, a \texttt{TGraphAsymmErrors} object is
written for each dependent variable.  If the data has finite bin widths, then
also separate \texttt{TH1F} objects are written for the central value of the
data points and each of the uncertainties.  If there is more than one
independent variable, the appropriate ROOT object (\texttt{TH2F} or
\texttt{TH3F}) is chosen instead of a \texttt{TH1F} object.
\item [\href{https://yoda.hepforge.org}{YODA}] The data analysis classes used
in the \href{https://rivet.hepforge.org}{Rivet} toolkit~\cite{Buckley:2010ar}.
 Again, the appropriate YODA object (\texttt{Scatter1D}, \texttt{Scatter2D},
\texttt{Scatter3D}) is written according to the number of independent variables
in a table.
\end{description}
The HEPData DOIs are written to the various output formats to allow a user to
later track the origin of the downloaded data points.  A further package,
\href{https://github.com/HEPData/hepdata-converter-ws}{\texttt{hepdata-converter-ws}},
runs the converter as a web service, which is deployed inside a
\href{https://www.docker.com}{Docker} container including dependencies such as
the \href{https://root.cern.ch}{ROOT} and
\href{https://yoda.hepforge.org}{YODA} packages.  Data can be accessed from a
user's script via predictable URLs, for example,
\begin{verbatim}
https://hepdata.net/record/ins1283842?format=yaml&table=Table1&version=1
\end{verbatim}
for \verb+format={csv,json,root,yaml,yoda}+.  Omitting the table name gives
all tables (in a \texttt{.tar.gz} file unless JSON) and omitting the version
number gives the latest version.  An option \texttt{light=True} for the JSON
format of a whole submission omits the data tables.

\section{Submission}

\begin{figure}
  \begin{center}
    \includegraphics[width=\textwidth]{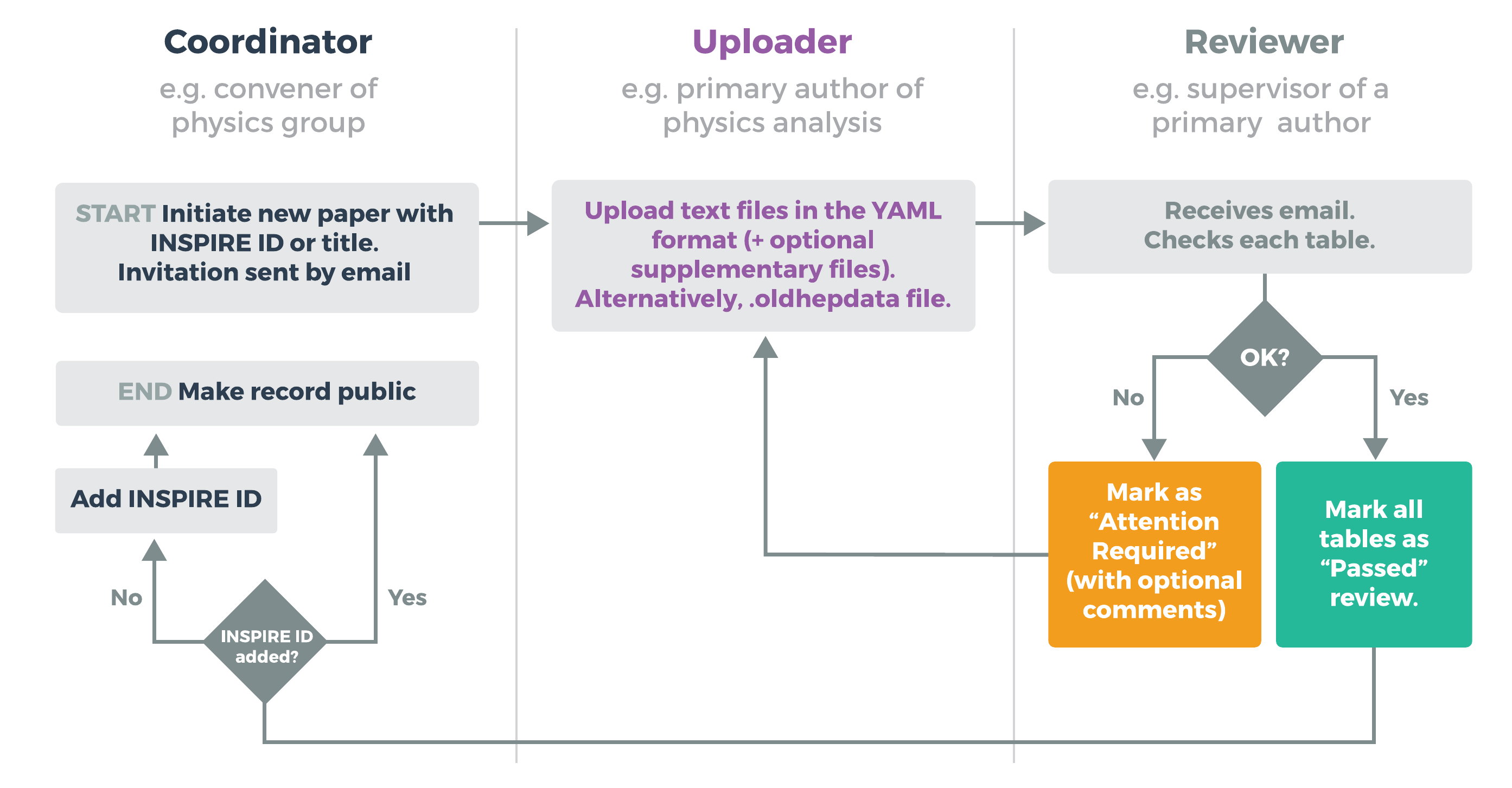}
  \end{center}
  \caption{\label{fig:submission}Submission flowchart for the new HEPData site.}
\end{figure}

\begin{figure}
  \begin{center}
    \frame{\includegraphics[width=\textwidth]{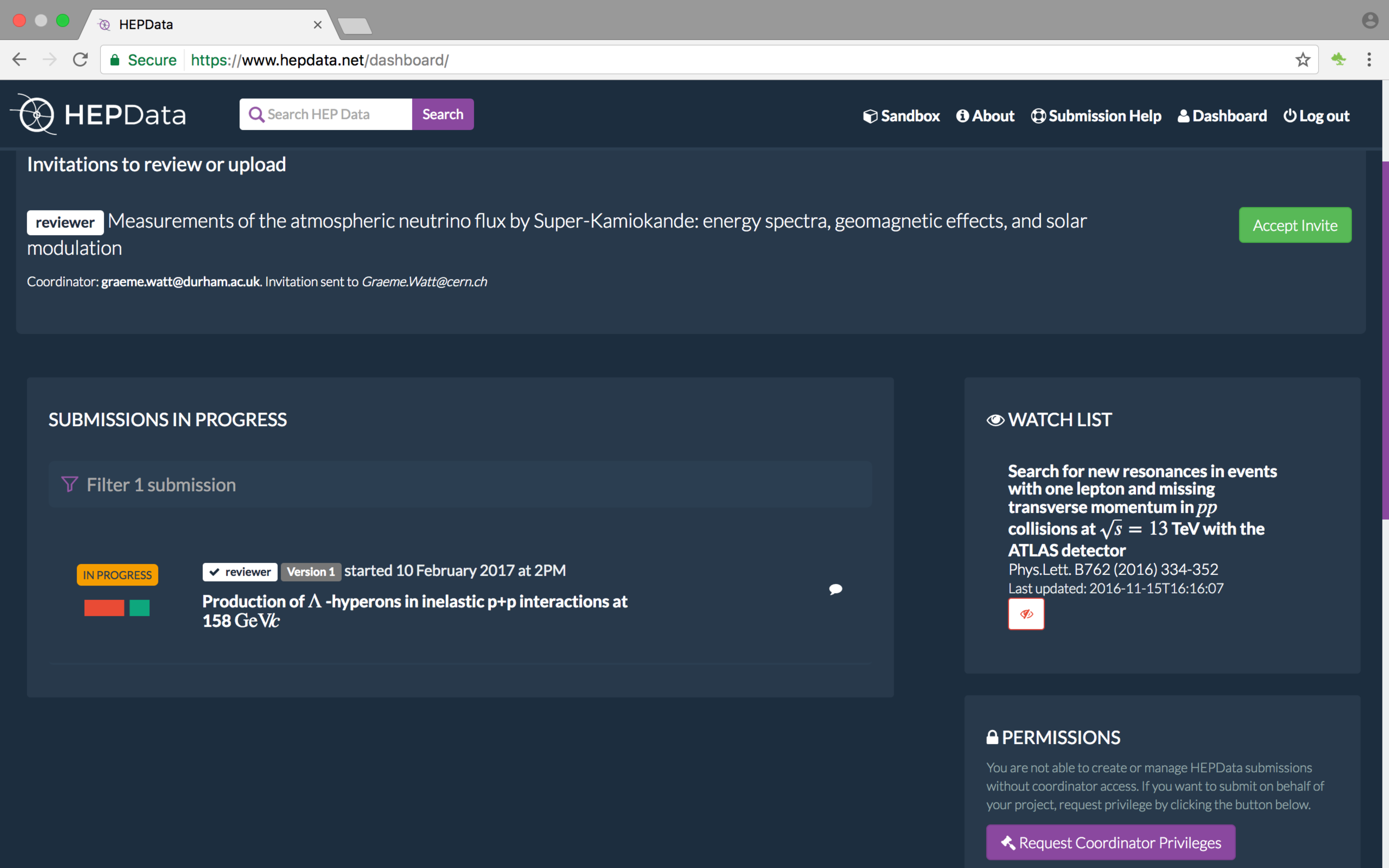}}
  \end{center}
  \caption{\label{fig:dashboard}Screenshot of the HEPData Dashboard for a user.}
\end{figure}

The new HEPData submission flowchart is shown in figure~\ref{fig:submission}.
The procedure has evolved from the workflow successfully used for external
submissions on the old HepData site for more than two years, with an average
external submission rate of around 12 papers/month, primarily from the four
main LHC experiments (ALICE, ATLAS, CMS, LHCb).  We define three different
roles:
\begin{description}
\item [Coordinator] Typically a fairly senior person within an experimental
collaboration, such as a convener of a physics group within ATLAS or CMS, who
initiates the submission and is responsible for its final approval on behalf of
the collaboration.
\item [Uploader] A primary author of a particular physics analysis, perhaps a
Ph.D.~student, who prepares and uploads the submission in one of the supported
standard input formats.
\item [Reviewer] A more senior person familiar with the particular physics
analysis, perhaps the supervisor of the primary author.  The Coordinator can
also choose to act as the Reviewer.
\end{description}
All current submissions are available for a Coordinator, Uploader, or Reviewer
to see in their \emph{Dashboard} (\url{https://hepdata.net/dashboard}); see
figure~\ref{fig:dashboard}.
Prospective Coordinators should select the ``Request Coordinator Privileges''
option from their Dashboard and enter the name of the experiment/group that
they wish to coordinate.  The request requires approval from the HEPData Admin
staff.  A list of current Coordinators can be seen at
\url{https://hepdata.net/permissions/coordinators}.  A logged-in user can
``Watch'' individual records to be notified by email of future revisions and a
user's ``Watch List'' is also accessible from their Dashboard; see
figure~\ref{fig:dashboard}.  An ``Ask a Question'' widget on each record allows
any logged-in user to send an email to the submission participants (or to the
Admin for migrated records) in order to point out mistakes or ask for
clarifications.

The Coordinator initiates a new submission by clicking the ``Submit'' button
after logging in, then entering either the Inspire
(\url{http://inspirehep.net}) record number of the corresponding publication
(assigned automatically after the paper appears on the
\href{https://arxiv.org/}{arXiv}) or a provisional paper title.  The
Coordinator assigns an Uploader and Reviewer, then an email is sent to the
designated Uploader with a link which gives them the privileges to upload a
data submission.  Once the Uploader thinks their submission is ready for
review, they should click the ``Notify Reviewer'' button on the record.  An
email is then sent to the designated Reviewer with a link which assigns them
the appropriate privileges.  The Reviewer marks each table as either
``Passed'' review or ``Attention Required'' via a widget displayed next to each
table, and messages can be entered to provide feedback to the Uploader.  The
Uploader can reply to these messages (which are also sent by email) via the
widget before uploading a revised submission with corrections.  Plots displayed
automatically beside each table help to find mistakes in numerical data points.
 The Reviewer needs to mark each table as having ``Passed'' review before the
Coordinator can ``Finalise'' the submission from their Dashboard.  The entire
submission will then be published and made searchable in HEPData.  It will also
appear on the \href{https://www.hepdata.net}{\texttt{hepdata.net}} homepage
under ``Recently Updated Submissions'' and a Tweet will be posted automatically
to the \href{https://twitter.com/HEPData}{@HEPData} Twitter account.  Further
explanation of the submission steps can be found at
\url{https://hepdata.net/submission} and in some independent
documentation~\cite{matteo_bonanomi_2016_197109}.

In contrast to the old HepData system, a submission can now be initiated and
reviewed without knowing the Inspire record number of the corresponding paper.
 This means that a HEPData record can be prepared simultaneously with the
corresponding paper.  The HEPData record can then be finalised soon after the
paper appears on the arXiv generating an Inspire entry.  Paper metadata is now
pulled directly from Inspire, therefore it does not need to be included
explicitly in the data input file, as was the case with the old HepData system.
 Assignment of DOIs is made when a HEPData record is initiated, although the 
DOIs are only minted when the record is finalised, meaning that the HEPData
record DOI can be cited in the first arXiv version of the corresponding
publication.

The primary submission format consists of a \texttt{submission.yaml} file
containing metadata together with a YAML (or JSON) data file for each table
containing the independent and dependent variables (see
\url{http://github.com/HEPData/hepdata-submission}).  In addition, there might
be some auxiliary files associated with either the whole submission or
individual tables.  All these files should be uploaded in a single archive
(\texttt{.zip}, \texttt{.tar}, \texttt{.tar.gz}).  However, if there are no
auxiliary files, the upload system also accepts a single YAML file containing
both metadata and data.  This format was used for migration and can be obtained
by appending ``\texttt{/yaml}'' to any of the old HepData record URLs.  To ease
the transition to the YAML format, the new submission system also accepts the
old ``input'' format in a single text file with extension \texttt{.oldhepdata},
which will be automatically converted by the
\href{https://github.com/HEPData/hepdata-converter}{\texttt{hepdata-converter}}
to the new YAML format.

A \emph{Sandbox} (\url{https://hepdata.net/record/sandbox}) allows any
logged-in user to upload a submission without any special permissions.  A
Sandbox record has a persistent URL (containing a 10-digit identifier) that can
be shared to allow access by anyone.  The Sandbox is therefore convenient for
testing uploads and sharing access to records that should not be made
searchable through the main HEPData site.  A Sandbox record can be removed (by
the user who created it) when it is no longer required.

\section{Future plans}

While HEPData has so far only been used for data associated with experimental
particle physics papers, it could easily be used to store numerical values of
theoretical predictions and related material from particle physics
phenomenology papers, without any necessary changes to the software or
submission workflow.  There is potential to store low-energy data from nuclear,
atomic, and medical physics, relevant for validation of the
Geant4~(\url{http://geant4.cern.ch}) detector simulation toolkit, but further
software development may first be needed to support keywords specific to the
low-energy data and to support creation of records where the associated
publications do not appear in the Inspire HEP literature database.

In future we plan to support a
\href{https://github.com/lukasheinrich/hepdata-rootcnv}{mixed YAML/ROOT input format}
where metadata is
provided in YAML files (as before), but numerical values are extracted from
ROOT objects and converted to the standard YAML format.
 HistFactory~\cite{Cranmer:1456844} is a framework used in many ATLAS studies
for statistical analysis (such as determining exclusion contours).  It encodes
the full likelihood (including systematic uncertainties) of a measurement
using semantic XML and histograms stored in ROOT files.  Some
\href{https://github.com/lukasheinrich/histfactory-cnvtools}{preliminary work}
has been done to extract HEPData tables in the standard YAML format directly
from a HistFactory configuration.  Furthermore, work has begun on expanding the
set of natively supported data types beyond a simple table to allow for richer
datasets such as HistFactory configurations or simplified
likelihoods~\cite{CMS:2242860}.  The archival of such likelihood data
in a lossless format could then be used by various reinterpretation packages.

\section{Summary}

The software underlying the Durham High Energy Physics database (HEPData) has
been completely rewritten over the last two years, predominantly in the Python
and JavaScript programming languages, as an overlay on the
\href{http://inveniosoftware.org}{Invenio}~v3 digital library framework, but
with a very large degree of customisation.  The new site
(\url{https://hepdata.net}) is now hosted at CERN on the OpenStack
infrastructure, but still managed remotely from Durham.  The transition from
the old site (\url{http://hepdata.cedar.ac.uk}) has effectively been completed,
with all data records being migrated to the new site.  The new submission
system has successfully been used for external data submissions from January
2017 onwards.

In conclusion, the new HEPData site provides a state-of-the-art web platform
for particle physicists to make their data \emph{Findable}, \emph{Accessible},
\emph{Interoperable}, and \emph{Reusable} according to the FAIR principles (see
\url{https://www.force11.org/group/fairgroup/fairprinciples}).

\ack
HEPData is funded by a
\href{http://gtr.rcuk.ac.uk/projects?ref=ST/N000315/1}{grant} from the UK
Science and Technology Facilities Council.  The DOI minting originates from the
THOR project, funded by the European Commission under the Horizon 2020
programme.  We are indebted to Mike Whalley for his dedicated 34 years of
service as Database Manager for previous incarnations of the HEPData project,
and for his assistance in migrating the data to the new platform.  We thank
Alicia Boya Garc\'{i}a, Kyle Cranmer, S\"unje Dallmeier-Tiessen, Frank Krauss,
Salvatore Mele, Laura Rueda, Jan Stypka and Micha\l{} Szostak for their various
contributions during the redevelopment process.

\section*{References}
\bibliography{chep2016_hepdata}
\bibliographystyle{iopart-num}

\end{document}